\documentstyle[12pt, epsfig]{article}

\def\be{\begin{equation}}
\def\ee{\end{equation}}
\def\bea{\begin{eqnarray}}
\def\eea{\end{eqnarray}}

\def\hpiNN{{h_{\pi NN}^{(1)}}}

\def\mbfa{{\mathbf a}}
\def\mbfA{{\mathbf A}}
\def\mbfI{{\mathbf I}}
\def\mbfj{{\mathbf j}}
\def\mbfJ{{\mathbf J}}

\def\mbfl{{\mathbf l}}
\def\mbfp{{\mathbf p}}

\def\mbfr{{\mathbf r}}

\def\mbfx{{\mathbf x}}

\begin{document}
\renewcommand{\thefootnote}{\#\arabic{footnote}}
\setcounter{footnote}{0}

\vskip 0.2cm

\begin{center}
{\Large \bf The Anapole Moment of the Deuteron with 
the Argonne $v18$ Nucleon-Nucleon Interaction Model}
\vskip 1.5cm
{\large Chang Ho Hyun}
$^{a,}$\footnote{e-mail: hch@color.skku.ac.kr}
and
{\large Bertrand Desplanques}
$^{b,}$\footnote{e-mail: desplanq@isn.in2p3.fr}

\vskip 0.1cm 
$^a${\it BK21 Physics Research Division,
Sung Kyun Kwan University,
Suwon 440-746, Korea} \\ 
$^b${\it Institut des Sciences Nucl\'{e}aires
(UMR CNRS/IN2P3-UJF),\\
F-38026 Grenoble Cedex, France}\\

\end{center}

\vskip 1cm

\centerline{\bf Abstract} \vskip 0.1cm

We calculate the deuteron anapole moment with 
the wave functions obtained from the Argonne $v18$
nucleon-nucleon interaction model. 
The anapole moment operators are considered at 
the leading order.
To minimize the uncertainty due to a lack of current conservation,
we calculate the matrix element of the anapole 
moment from the original definition.
In virtue of accurate wave functions, we can obtain 
a more precise value of the deuteron anapole moment 
which contains less uncertainty 
than the former works.
We obtain a result reduced by more than 25\% in the 
magnitude of the deuteron anapole moment. 
The reduction of individual nuclear contributions is much 
more important however, varying from a factor 2 for the 
spin part to a factor 4 for the convection and 
associated two-body currents.

%

\newpage

\section{Introduction}

After Zel'dovich introduced the concept of the anapole moment (AM)
\cite{zel}, a first non-zero measurement was reported quite 
recently in the $^{133}$Cs atom \cite{wood}.
Isolation of the effect is difficult 
because the AM has a small contribution 
compared to the leading order $Z^0$ exchange 
in atomic physics.
However, the part that involves the nuclear spin is 
suppressed by a factor $1 - 4 \sin\theta^2_w\ (\simeq 0.08)$.
Thus in some cases, the AM which is a higher-order effect in 
electro-weak interactions can be comparable with
the spin-dependent $Z^0$ contribution.
Flambaum and Khriplovich showed that
the AM of a heavy nucleus is proportional 
to $A^{2/3}$ \cite{fk80},
from which one can deduce the dominance of
the AM over $Z^0$ exchange or radiative corrections
for large $A$.
Since then, calculations of the AM 
of heavy nuclei have been the object of the major interest
in the theoretical works of the domain
\cite{fk80, fks84, dkt94, dt97, hlr01}.

The dominant contribution to the AM 
of a light system, namely the deuteron, was pointed out
by one of the author \cite{des81}.
The spin current which stems from the 
anomalous magnetic moment of the nucleon 
was expected to give a few times larger contribution
than the convection current or exchange currents
and it was verified in several papers
\cite{hh81, ssnpa98, kknpa00, savagenpa01}.
As a result, the contribution of the AM 
becomes similar to the radiative corrections.
Khriplovich and Korkin (KK in short) calculated
the AM of the deuteron analytically with the 
zero-range-approximated (ZRA) wave functions 
\cite{kknpa00}.
Their result of the spin-current term, which is the most 
dominant contribution to the AM of the deuteron,
has exactly the same form as the one obtained 
from the framework of the effective field theory 
by Savage and Springer (SS in short) \cite{ssnpa98}. 
The error of the result, which may 
be mainly from the simple wave functions, is 
estimated to be about 20\%.
However, in recent calculations of the asymmetry in the 
$\vec{n} + p \rightarrow d + \gamma$, we showed that
the ZRA (or the effective field theory \cite{kssw99})
result exceeds the ones obtained with a few
phenomenological wave functions by more than  50\% 
\cite{desplb01, hpmplb01}.
With this observation, it may be possible that the 
error of the deuteron AM
with ZRA wave functions can be 
larger than 20\%.
With the purpose to minimize theoretical uncertainties,
we calculate the deuteron AM with the wave 
functions obtained from the Argonne $v18$ potentials
\cite{wssprc95}.

The low momentum transfer which characterizes the process 
makes it possible to treat the problem with
heavy-baryon-chiral-perturbation theory (HBChPT).
With the counting rules of HBChPT, 
one can obtain the transition operators 
order-by-order in a well-defined way. 
The error of a calculation can be systematically
estimated from the 
higher orders that are not taken into account
in the calculation.
Since well-ordered operators are evaluated with
very accurate phenomenological wave functions,
we expect that the uncertainties will be the least 
among the theoretical calculations.









\section{The Anapole Moment}

The anapole moment of a system is obtained from the expansion 
of the vector-potential
\begin{eqnarray}
{\rm A}_i (\mbfx) = \int d \mbfx' 
\frac{{\rm j}_i(\mbfx')}{|\mbfx - \mbfx'|}
\end{eqnarray}
in a series in $x^{-1}$ ($x \equiv |\mbfx|$).
The quantity $\mbfj$ is the matrix element of the current density
operator for given initial and final states
\begin{eqnarray}
\mbfj = <\psi_f |\, \hat{\mbfj}\, | \psi_i>.
\end{eqnarray}
The zeroth order term vanishes since there is no net current
and the first-order term gives the vector potential of the 
magnetic dipole.
The second order can be separated 
into a magnetic quadrupole term and the anapole term.
After some algebra, one obtains the following form of the 
anapole term
\footnote{While we have the AM in fm$^2$, the Seattle group favors
a dimensionless definition which can be obtained by 
multiplying ours by $m^2_N/4\pi$.}
\begin{eqnarray}
\mbfa \equiv \frac{2 \pi}{3} \int d\mbfx \;
\mbfx \times (\mbfx \times \mbfj(\mbfx))
\label{eq:orig-am}
\end{eqnarray}
with which the vector potential of the anapole moment
reads
\begin{eqnarray}
\mbfA_{\rm\small anapole}(\mbfx) = 
\frac{1}{4 \pi} \left( - \mbfa \; \nabla^2 \frac{1}{x}
+ \mbfa \cdot \nabla \; \nabla \frac{1}{x} \right).
\end{eqnarray}
The second term of $\mbfA_{\rm\small anapole}$ can be removed
by a suitable choice of gauge and 
the resultant anapole vector potential
takes the form 
\begin{eqnarray}
\mbfA_{\rm\small anapole}(\mbfx) = \mbfa\; \delta^{(3)}(\mbfx).
\end{eqnarray}
If current conservation is satisfied, 
Eq. (\ref{eq:orig-am}) can also be written as
\begin{eqnarray}
\mbfa = - \pi \int d\mbfx \; x^2 \; \mbfj(\mbfx).
\label{eq:am-def}
\end{eqnarray}
In many calculations of the nuclear AM, Eq. (\ref{eq:am-def})
was adopted as the working definition of the AM.
However, as shown in \cite{hlr01}, the contribution of
a current term depends strongly on this definition  
while the total result may be sensitive on fulfilling 
the current-conservation constraint. 
The situation becomes more uncertain for
the exchange currents or higher order terms.
Thus in order to avoid possible large effect on the result 
due to lack of current conservation, 
we preferentially adopt in our calculations the original definition, 
Eq. (\ref{eq:orig-am}). We nevertheless stress that we 
include in our work a minimal set of two-body currents 
ensuring current conservation in relation with the 
parity-non-conserving (PNC) one-pion exchange interaction. 
This one (together with the PC part) turns out to be essential
in ensuring the approximate equivalence of Eq. (6) with Eq. (3),
as will be seen at the end of the paper.


\section{Operators and Wave Function}

\begin{figure}[h]
\begin{center}
\epsfig{file=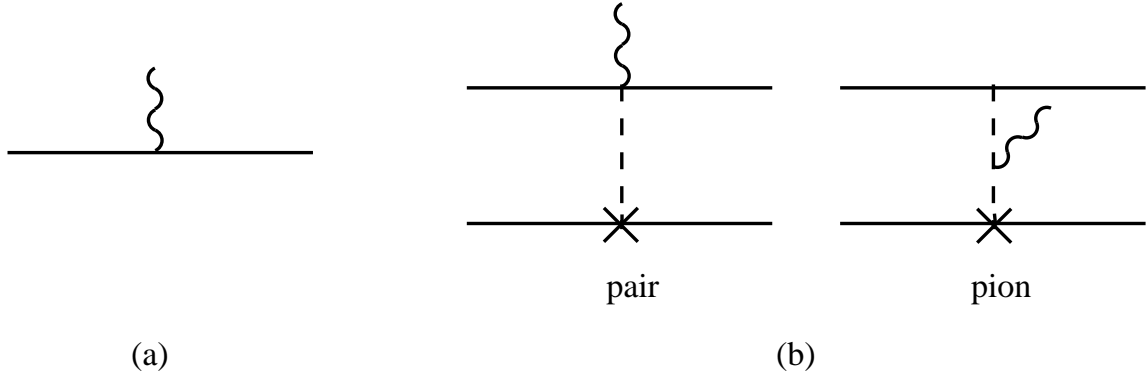, width = 6in}
\caption{Diagrams representing one and two-body electromagnetic contributions.}
\end{center}
\end{figure}

The diagrams considered are shown in Fig. 1.
Leading one-body currents (Fig. 1-(a))
are composed of spin and convection currents 
and PNC exchange contributions are in Fig. 1-(b).
The leading PNC vertex marked with ${\mathbf \times}$
reads \cite{ddh80}
\begin{eqnarray}
{\cal L}^{(1)}_{\pi NN} = - \frac{\hpiNN}{\sqrt{2}}
N^\dagger (\vec{\tau} \times \vec{\pi})^z N.
\end{eqnarray}
The current density operators of each diagram are
\begin{eqnarray}
\hat{\mbfj}_{\rm\small spin}(\mbfx) &=& e \sum^2_{i = 1}
\frac{\mu^i_N}{2 m_N} \nabla_\mbfx \times
(\vec{\sigma}_i\ \delta^{(3)}(\mbfx - \mbfr_i)), \\
\hat{\mbfj}_{\rm\small conv}(\mbfx) &=& e \sum^2_{i = 1}
\frac{1 + \tau^z_i}{4 m_N} \left\{
\mbfp_i, \delta^{(3)}(\mbfx - \mbfr_i) \right\}, \\
\hat{\mbfj}_{\rm\small pair}(\mbfx) &=& 
- e\, \frac{g_A \hpiNN}{2 \sqrt{2} f_\pi} 
\left(\vec{\tau}_1 \cdot \vec{\tau}_2 - \tau^z_1 \tau^z_2\right)
y_0(r_{12})
\sum^2_{i = 1} \vec{\sigma}_i\ \delta^{(3)}(\mbfx - \mbfr_i), \\
\hat{\mbfj}_{\rm\small pion}(\mbfx) &=& 
- e\, \frac{g_A \hpiNN}{2 \sqrt{2} f_\pi}
\left(\vec{\tau}_1 \cdot \vec{\tau}_2 - \tau^z_1 \tau^z_2\right)
\left(\vec{\sigma}_1 \cdot \vec{\partial}_1 
- \vec{\sigma}_2 \cdot \vec{\partial}_2 \right) 
(\vec{\partial}_1 - \vec{\partial}_2)
y_0(r_{1x})\, y_0(r_{2x})
\end{eqnarray}
where $r_{12} \equiv |\mbfr_1 - \mbfr_2|$,
$r_{i x} \equiv |\mbfr_i - \mbfx|$ 
and
\[
y_0(r) \equiv \frac{{\rm e}^{-m_\pi r}}{4 \pi r}.
\]
$\mu^i_N$ is defined as
\[ \mu^i_N \equiv \frac{1}{2}\left(
\mu_S + \tau^z_i \mu_V \right)
\]
with $\mu_S = 0.88$ and $\mu_V = 4.71$.
The PNC interaction of the proton and the neutron
generates parity-odd components in the deuteron wave function.
In the context of the meson-exchange picture, 
the PNC interaction is mediated 
by $\pi$, $\rho$, $\omega$ and heavier mesons.
PNC components in the wave function 
can be obtained by solving the Schr\"{o}dinger 
equation with the PNC potentials given in \cite{ddh80}.
In a low energy process, it is believed that the pion 
exchange will dominate PNC interactions when its contribution 
is not forbiden by some selection rule. 
The pion-exchange PNC potential reads
\begin{eqnarray}
V^{1\pi}_{\rm\small pnc}(\mbfr) = 
\frac{g_A \hpiNN}{\sqrt{2} \, f_\pi} (\vec{\tau}_1 \times \vec{\tau}_2)^z \; 
\mbfI \cdot \hat{\mbfr} \; \frac{d}{dr} y_0(r)
\label{eq:pncpot}
\end{eqnarray}
where $\mbfr \equiv \mbfr_p - \mbfr_n$, $r \equiv |\mbfr|$
and $\mbfI \equiv \frac{1}{2}(\vec{\sigma}_p + \vec{\sigma}_n)$. 
The constants, $g_A$ and $f_\pi$ are given the values 1.267 and 92.4 MeV
respectively.
For the deuteron,
the PNC potential, Eq. (\ref{eq:pncpot}) gives rise to 
a parity-odd $^3$P$_1$ component. 
We write the parity admixed wave function as
\begin{eqnarray}
\psi_d(\mbfr) &=& \frac{1}{\sqrt{4 \pi} r}
\left[ 
\left( u(r) + S_{12}(\hat{\mbfr}) \frac{w(r)}{\sqrt{8}} \right) \zeta_{00}
- i \hpiNN 
\sqrt{\frac{3}{2}} \; \mbfI \cdot \hat{\mbfr} \; v(r) \; \zeta_{10} \right]
\chi_{1 J_z}
\end{eqnarray}
where $S_{12}(\hat{\mbfr}) \equiv 3 \; \vec{\sigma}_1 \cdot \hat{\mbfr} \;
\vec{\sigma}_2 \cdot \hat{\mbfr} - \vec{\sigma}_1 \cdot \vec{\sigma}_2$
and $\chi$ and $\zeta$ represent a spinor and an isospinor, respectively.
The quantities $u(r),\ w(r)$ and $v(r)$ are obtained by solving the 
Schr\"{o}dinger
equation and one can calculate the matrix elements of the current density 
operators, Eqs. (8) -- (11), with the obtained solutions. 
After that, the calculation of the anapole moment with Eq. (3) is 
straightforward.

\section{Results}

The anapole moment, Eq. (3), reads for each term
\begin{eqnarray}
\mbfa_{\rm\small spin} &=& - \mu_V
\sqrt{\frac{1}{6}} \frac{\pi}{m_N} 
\int dr\, r\, v(r)\, \left(u(r)\, - \, \sqrt{2} w(r)\right) \
e\, \mbfI\, \hpiNN, \\
\mbfa_{\rm\small conv} &=& \ \ \ \
\frac{1}{3}
\sqrt{\frac{1}{6}} \frac{\pi}{m_N} 
\int dr\, r\, v(r)\, \left(u(r)\, - \, \sqrt{2} w(r)\right) \
e\, \mbfI\, \hpiNN, \\
\mbfa_{\rm\small pair} &=& - 
\frac{\sqrt{2} \pi g_A}{9 f_\pi}
\int dr\, r^2\, y_0(r)\, 
\left(u(r) + \frac{1}{\sqrt{2}} w(r)\right)
\left(u(r) - \sqrt{2} w(r) \right)
e\, \mbfI\, \hpiNN, \\
\mbfa_{\rm\small pion} &=& \ \
\frac{\sqrt{2} \pi g_A}{3 f_\pi m_\pi}
\int dr\, r\, y_0(r)
\left( u(r) + \frac{1}{\sqrt{2}} w(r) \right)
\times \nonumber \\
& & \ \ \ \ \ \ \ \ \ \ \ \ \ \ \ \ \ \ \ \  
\left[ u(r) \left(1 - \frac{1}{3} m_\pi r \right)
- \frac{1}{\sqrt{2}} w(r)
\left(1 + \frac{1}{3} m_\pi r \right) \right]
e\, \mbfI\, \hpiNN.
\end{eqnarray}
Numerical results are
\begin{eqnarray}
\mbfa_{\rm\small spin} &=&   - 0.531\, e \mbfI\, \hpiNN, \\
\mbfa_{\rm\small conv} &=& \ \ \, 0.038\, e \mbfI\, \hpiNN, \label{eq:conv}\\
\mbfa_{\rm\small pair} &=&   - 0.026\, e \mbfI\, \hpiNN, \label{eq:pair}\\
\mbfa_{\rm\small pion} &=& \ \ \, 0.027\, e \mbfI\, \hpiNN \label{eq:pion}
\end{eqnarray}
where all the values are in the fm$^2$ unit.

\begin{figure}[tbp]
\begin{center}
\epsfig{file=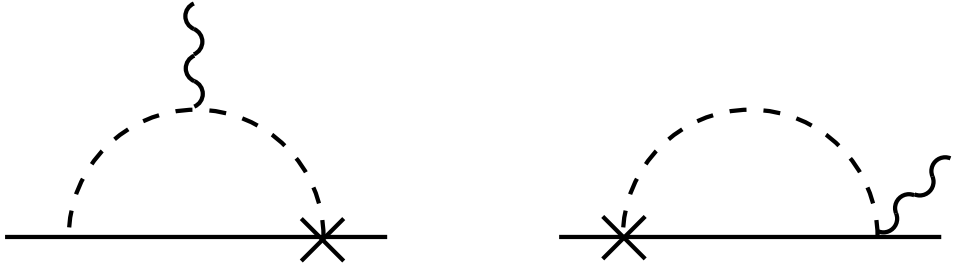, width= 5in}
\caption{The diagrams of the leading order
nucleonic anapole moment. Solid, dashed and wavy
lines represent the nucleon, pion and photons, 
respectively.}
\end{center}
\end{figure}

In the HBChPT,
one can specify the magnitude of a diagram in terms of
the power ($\nu$) of the momentum transfer ($Q$)
by the rule
\begin{eqnarray}
\nu = 2 L - 2 (C - 1) - 1 + \sum_i \nu_i
\end{eqnarray}
where
\begin{eqnarray}
\nu_i = d_i + \frac{n_i}{2} + e_i - 2.
\end{eqnarray}
$L$ is the number of loops, $C$ is the number 
of disconnected lines, $d_i$, $n_i$ and $e_i$
are the numbers of derivatives, nucleon lines
and external gauge fields at the vertex $i$,
respectively. All the strong and electro-magnetic
vertices satisfy 
\begin{eqnarray}
\nu_i \geq 0
\end{eqnarray}
but the PNC vertex, Eq. (7) has $\nu_i = -1$.
One can easily verify that the diagrams in Fig. 1
have $\nu = -2$. 
Nucleon anapole diagrams shown in Fig. 2 also have
this order. In order to be consistent with the 
order counting of the effective field theory,
the nucleonic terms also should be included.

This term was already calculated in several works
\cite{hhmprl89, ksnpa93, zphmprd00, mkplb00}.
Since the results are consistent in both magnitude
and sign, we do not repeat its calculation but just 
adopt the result here. The leading order nucleonic anapole,
taking into account the deuteron D-state probability $P_D$, reads
\begin{eqnarray}
\mbfa_{\rm\small N} &=& - 
\frac{g_A}{6 \sqrt{2} f_\pi m_\pi}\;(1-\frac{3}{2}\,P_D)\;
e\, \mbfI\, \hpiNN \nonumber \\
&=& - 0.417 \, e\, \mbfI\, \hpiNN.
\end{eqnarray}
Then the anapole moment of the deuteron at the leading order
is 
\begin{eqnarray}
\mbfa_{\rm\small d} &=& 
\mbfa_{\rm\small spin} + \mbfa_{\rm\small conv}
+ \mbfa_{\rm\small pair} + \mbfa_{\rm\small pion}
+ \mbfa_{\rm\small N} \nonumber \\
&=& - (0.531 - 0.038 + 0.026 - 0.027 + 0.417)
\, e\, \mbfI\, \hpiNN \nonumber \\
&=& - 0.909 \, e\, \mbfI\, \hpiNN.
\end{eqnarray}

\section{Discussions}

As expected, the spin term is the most dominant one
among the contributions from the currents.
The contributions from the convection, pair and pion terms 
are much smaller while the two last ones strongly cancel.
The sum of these three contributions is only
about 7\% of the spin term.
However the nucleonic term is close to the spin term,
which is quite contradictory to the former results
of the deuteron anapole moment.

In order to illustrate this feature clearly, let
us compare our result with the ones 
by SS \cite{ssnpa98} and KK \cite{kknpa00}.
Firstly, SS's result is
\begin{eqnarray}
\mbfa^{\rm\small SS}_{\rm\small d} &=&
\mbfa^{\rm\small SS}_{\rm\small spin} +
\mbfa^{\rm\small SS}_{\rm\small PE} +
\mbfa_{\rm\small N} \nonumber \\
&=& - (1.03 - 0.18 + 0.46) \, e\, \mbfI\, \hpiNN
= - 1.31 \, e\, \mbfI\, \hpiNN,
\label{eq:amss}
\end{eqnarray}
where the pion-exchange term (PE) corresponds to the sum
of the convection, pair and pion terms in our calculation.
We have verified in a separate calculation 
that SS's result can be obtained with the ZRA wave function
\cite{desnote02}.
Secondly, KK's result reads
\begin{eqnarray}
\mbfa^{\rm\small KK}_{\rm\small d} &=&
\mbfa^{\rm\small KK}_{\rm\small spin} +
\mbfa^{\rm\small KK}_{\rm\small orb} +
\mbfa^{\rm\small KK}_{\rm\small N} \nonumber \\
&=& - (1.03 - 0.07 + 0.21 ) \, e\, \mbfI\, \hpiNN
= - 1.17 \, e\, \mbfI\, \hpiNN.
\label{eq:amkk}
\end{eqnarray}
Since KK also used the ZRA wave function, their spin term 
is equal to SS's one. However there are substantial 
differences in the remaining terms. In the following,
we successively discuss these different contributions.

The biggest quantitative difference of our result 
with the above ones comes from the spin term. 
The $Av18$ result is roughly half of the ZRA value.
In \cite{kknpa00}, KK argued that the uncertainty of the
ZRA wave function is about 20\%. 
Taking this uncertainty into consideration,
the spin term can be reduced to $-0.82 \, e\, \mbfI\, \hpiNN$
which is still larger than our result by about 55\%.
The discrepancy of our result with the KK (as well as SS)
estimate  is understood by noting that our calculation 
incorporates the effect of a short-range repulsion 
in S-states as well as the known repulsive character 
of the NN interaction in the $^3P_1$ channel.
We also note that the ZRA wave function contains
only a central component, $u(r)$, while our calculation
includes a tensor one, $w(r)$ too. Contrary to the PNC 
asymmetry in n-p radiative capture \cite{desplb01, hpmplb01},
this component produces a partly destructive interference. 
The contribution of each part is 
\begin{eqnarray}
\mbfa^{\rm\small cen}_{\rm\small spin} 
&=& -0.724 \, e\, \mbfI\, \hpiNN \\
\mbfa^{\rm\small ten}_{\rm\small spin}
&=& \ \ \, 0.193 \, e\, \mbfI\, \hpiNN.
\end{eqnarray}
In some cases, the contribution of the tensor part in the 
wave function is assumed to be small and its contribution
is neglected. However in the deuteron anapole moment,
this contribution is non-negligible and its effect on the
magnitude of the anapole moment is substantial.

We now turn to the second contribution 
in Eqs. (\ref{eq:amss}, \ref{eq:amkk}),
which, among other contributions, 
involves the convection current one.
There too, a large suppression of our results
is observed but the effect is more drastic for
the PE term in SS's result than for the orbital
one in KK's result. 
In SS's result, PE term's contribution is about 15\% of the
total magnitude, but in our result it is only 4\% 
of the total value.  Then, the question arises why the
PE term in SS does not coincide with the orbital term in KK.

It can be shown that the anapole moment of the convection term with
the definition Eq. (3) is equivalent to the orbital term
\begin{eqnarray}
\mbfa_{\rm\small conv} = \left<
- i \frac{\pi e}{12 m_N}\left[\mbfl^2,\ \mbfr \right]
\right>
\end{eqnarray}
where $\mbfl \equiv \mbfr \times \mbfp$ is the 
angular momentum operator in the center of mass frame.
The global factor of the sum of the spin and the 
convection term,
$\left(\mu_V - 1/3 \right)$ always appears regardless
of wave functions. This explains the relative ratio 
of the contributions due to the spin and convection currents
in KK's work as well as in ours. 

As mentioned above, KK's orbital term accounts for only 
the convection term. In KK's argument, they made use of the
fact that 
\begin{eqnarray}
\mbfJ_{\rm\small pair}(\mbfr) +
\mbfJ_{\rm\small pion}(\mbfr) \propto \mbfr.
\label{eq:coeq}
\end{eqnarray}
With the definition Eq. (3), it may be that
this term gives a zero contribution since
\begin{eqnarray} 
\mbfa \propto \mbfr \times (\mbfr \times \mbfJ).
\label{eq:wrongam}
\end{eqnarray}
We would like to point out that caution is required in
the above reasoning.
Firstly, the matrix elements that are relevant in the calculation
of the anapole moment involve the {\it current density} operator and
not the {\it current} operator. It can be easily shown
that the current operator which is obtained by integrating the 
current density operator with respect to the field point $\mbfx$
does satisfy Eq. (\ref{eq:coeq}). 
However the anapole moment should be
obtained from the double vector product of the current 
density operator and field point $\mbfx$, whose
result may not be like Eq. (\ref{eq:coeq}) in general.
Secondly, a rough substitution $\mbfx \rightarrow \mbfr$
in Eq. (3) gives Eq. (\ref{eq:wrongam}).
This substitution, even if one is working in the center of
mass frame and thus center of mass coordinate is discarded,
should be carefully derived from the evaluation of the matrix
elements with the two-body wave function.
A crude transformation of the coordinates may give wrong results.
As our result shows, a rigorous derivation with the 
{\it current density} operator gives a non-zero contribution
of the two-body currents.
We also checked that this derivation provides
coincidence with SS's result \cite{desnote02}. Curiously,
the extra contributions that the SS's work accounts for do
not seem to show up in our results. It turns out that the term
proportional to $1/m_{\pi}$ in Eq. (\ref{eq:pion}), a priori favored,
is suppressed with the ZRA value (numerical factor 0.07 instead 
of 0.18). It is reminded that in the zero-pion-mass limit, 
this term is cancelled by another one arising from the 
nucleonic term \cite{ssnpa98}. The other term (Eq. (\ref{eq:pair})
and part of Eq. (\ref{eq:pion})), which is less singular
in configuration space and has an opposite sign, 
is essentially unchanged. As a result, the sum 
of the two contributions almost vanishes in our work. 

Concerning the nucleonic term
(third one in Eqs. (\ref{eq:amss}, \ref{eq:amkk})), KK adopt
\begin{eqnarray}
\mbfa^{\rm\small KK}_{\rm\small N} =
- \frac{g_A}{6 \sqrt{2} f_\pi m_\pi}
\left(1 - \frac{6 m_\pi}{\pi m_N} \ln\frac{m_N}{m_\pi} \right)
\, e\, \mbfI\, \hpiNN.
\end{eqnarray}
The first term in the parenthesis coincides with the one 
we retained and the second 
term stems from the $1/m_N$ correction to the first term.
In the context of the counting rule of the effective field theory,
this $1/m_N$ term is classified in the higher order corrections
to the leading term. In order to be consistent with the 
strategy of the effective field theory, the $1/m_N$ corrections 
should be taken into account consistently, i.e. their correction
should be calculated not only for the nucleonic term
but also for current terms or any other types that have the same 
order. In that sense, KK's result contains the $1/m_N$ correction
partially. However, it is interesting to notice that the magnitude
of the $1/m_N$ correction to the nucleonic term amounts to 
a half of its leading value. This indicates that the higher order
corrections can play a critical role in the magnitude of the 
deuteron anapole moment. Their calculation should be done 
in the future.

In a separate work, we have observed that, while the spin current
satisfies gauge invariance by itself, convection, pair and 
pion currents are not gauge invariant solely and gauge invariance
is restored when the three terms are summed up in a minimal case 
\cite{desnote02}. In the general case, one should also include 
two-body currents related to the description of the NN strong interaction.
As a way to investigate how much gauge invariance is broken,
we compare the anapole moments calculated with Eq. (3)
and Eq. (6). Equation (3) is defined from the definition of the
anapole vector potential and Eq. (6) is derived from Eq. (3)
by imposing current conservation. Calculation of the 
anapole moment from Eq. (6) is straightforward too. 
The results are
\begin{eqnarray}
\mbfa^{\rm\small CC}_{\rm\small conv} &=&
\ \ \, 0.051\, e\, \mbfI\, \hpiNN , \label{eq:conv2} \\
\mbfa^{\rm\small CC}_{\rm\small pair} &=& 
-0.076\, e\, \mbfI\, \hpiNN , \label{eq:pair2}\\
\mbfa^{\rm\small CC}_{\rm\small pion} &=&
\ \ \, 0.045\, e\, \mbfI\, \hpiNN. \label{eq:pion2}
\end{eqnarray}
where the superscript CC implies the result with the 
assumption of current conservation. As expected, individual 
contributions differ from those in Eqs. (\ref{eq:conv}-\ref{eq:pion}),
but it is also noticed that the sum differs, 
$0.020\, e\, \mbfI\, \hpiNN$, 
instead of $0.039\, e\, \mbfI\, \hpiNN$. 
The first candidate to explain the discrepancy 
is the two-body current associated to the strong 
one-pion exchange interaction. The corresponding 
contribution, $0.018\, e\, \mbfI\, \hpiNN$, 
which should be added to the sum of contributions of Eqs.
(\ref{eq:conv2}-\ref{eq:pion2}), 
removes most of the difference. 
This one-pion exchange interaction
can also generate self-gauge invariant contributions 
that will affect equally the sum of contributions 
in Eqs. (\ref{eq:conv}-\ref{eq:pion}) 
and Eqs. (\ref{eq:conv2}-\ref{eq:pion2}).
Due to cancellations of pair and pion contributions,
this common contribution is found to be small
($-0.0007\, e\, \mbfI\, \hpiNN$).

Concluding, we calculated the anapole moment of the deuteron
with the wave functions obtained from the $Av18$ potential.
Its magnitude is reduced by more than 25\% from the previous ZRA 
results but is still comparable to the radiative corrections. 
If the contribution of the nucleon anapole moment is put aside,
the effect is much larger however, ranging from a factor 2 
for the spin contribution to a factor 4 for the contribution
of the convection current and associated two-body currents. 
We observed that gauge invariance for these last contributions,
hopefully smaller,  is a severe constraint and its fulfillment
is a non-trivial problem. In the analysis of other's work, 
we noticed that the corrections of higher orders can modify
significantly their estimate. The calculation of gauge-invariant 
higher-order corrections will be the future challenge.

\end{document}